\documentstyle[prl,twocolumn,aps]{revtex}
\begin{document}
\draft
\title{Statistical Origin of Quantum Mechanics}
\author{ G. Kaniadakis }
\address{ Dipartimento di Fisica and Istituto Nazionale di
 Fisica della Materia\\ Politecnico di Torino,
Corso Duca degli Abruzzi 24, 10129 Torino, Italy \\ Email address:
kaniadakis@polito.it}

\date{\today}
\maketitle

\begin {abstract}
The one particle quantum mechanics is considered in the frame of a
N-body classical kinetics in the phase space. Within this
framework, the scenario of a subquantum structure for the quantum
particle, emerges naturally, providing an ontological support to
the orthodox quantum mechanics. This approach to quantum
mechanics, constitutes a deductive and direct method which, in a
self-consistent scheme of a classical kinetics, allows us: i) to
obtain the probabilistic nature of the quantum description and to
interpret the wave function $\psi$ according to the Copenhagen
school; ii) to derive the quantum potential and then the
Schr\"odinger equation; iii) to calculate the values of the
physical observables as mean values of the associated quantum
operators; iv) to obtain the Heisenberg uncertainty principle.

\end {abstract}

\pacs{ PACS number(s): 03.65.-w, 03.65.Bz, 03.67.-a, 03.67.Hk,
03.67.Lx, 03.67.Dd }

\section{Introduction}

The problem still open of interpretation and derivation of quantum
mechanics in the frame of classical physics was first arisen in
the same year in which the Schr\"odinger equation first appeared.
Essentially two different approaches to this problem exist.

In the earliest, the hydrodynamic approach, introduced in 1926,
Madelung showed that the ansatz $\psi=\rho^{1/2}\exp(i {\rm
S}/\hbar)$ transforms the time dependent Schr\"odinger equation
into two hydrodynamic equations \cite{M}. The first one is the
evolution equation for $\rho$ and is a continuity equation while
the second is the evolution equation for $S$ and can be viewed as
an Euler equation or as an Hamilton-Jacobi equation with an extra
contribution to the energy corresponding to the presence of the
quantum potential. The quantum fluid obtained in this way starting
from the Schr\"odinger equation was studied extensively by various
authors \cite{Ho,HB,B,T1,T2,KK,F1,F2,F3,F4,F5,F6,F7,CW,TW}.
Takabayashi has shown that the quantum potential leads to a
particular form for the stress tensor of the quantum fluid
\cite{T1}. In the hydrodynamic approach which can be viewed
essentially as an indirect method that permits to interpret the
quantum mechanics, the particle takes the form of a highly
localized inhomogeneity that moves with the local fluid velocity
so that the quantum fluid appears as the medium which transport
the particle.

The second, the stochastic mechanics approach
\cite{F,N,S1,S2,S3,S4,S5,S6,S7,S8,S9,S10,S11,G}, is more ambitious
being a direct method which wants to derive the Schr\"odinger
equation in the frame of a particular classical dynamics and gives
an explicit representation of quantum mechanics in term of
classical probability densities for a particle undergoing a
brownian motion with a diffusion constant proportional to Planck's
constant and inversely proportional to particle mass. This quantum
brownian motion is nondissipative in contrast to the standard one
and is responsible for the disordering of the particle trajectory
and does not take energy from it. The dynamics of this quantum
brownian particle is described in the frame of the
Einstein-Smoluchowski theory which is an approximation of the
Ornestein-Uhlenbeck theory governing the standard brownian motion.
The main critics to the stochastic approach is that the quantum
brownian motion cannot be obtained from first principles and
appears built on specifically in order to produce the quantum
mechanics. In this way it seems that the stochastic approach moves
the problem to another level \cite{S/1,S/2,S/3,S/4,S/5}.

The idea that the quantum state corresponds to a classical
statistical ensemble is common in both the approaches above
described. This idea, so commonly accepted, we believe must to be
at the base of any attempt to derive and interpret the quantum
mechanics. Let's pose now the question: does it exist a rigorous
way to describe a time depending classical statistical ensemble
starting from some simple principle without making use of any
extra assumption in order to obtain the quantum mechanics? In
other words, is it possible to build a program in which the
microscopic motion, underlying the quantum mechanics, is described
by a rigorous dynamics different from the brownian one? The answer
to this question is affirmative.

In the present paper we will show that the quantum state
corresponds exactly to a subquantum statistical ensemble whose
time evolution is governed by a classical kinetics in the phase
space. The idea that the classical kinetics could be underlying
quantum mechanics is new and is proposed here the first time.

\section{Classical kinetics}

Let us consider a set of $N$ identical interacting particles in a
$n$-dimensional physical space ($n=1,2,3$) which in the following
we call {\it monads}. Each monad has a mass $\mu$ and obeys to the
laws of classical physics. We suppose that this set classical
particles constitutes a statistical ensemble described in the
$2n$-dimensional phase space by the distribution function $ f(t,
\mbox{\boldmath $x$} , \mbox{\boldmath $v$})$ with
\begin{equation}
\int f\, d^{n}v d^{n} x=N \ \ ,
\end{equation}
and postulate that the dynamics of this set of particles is
governed by the following kinetic equation:
\begin{equation}
\frac{\partial {f}}{\partial t}
 + \mbox{\boldmath $v$} \frac{\partial {f} }{\partial
  \mbox{\boldmath $x$}}
+ \frac{\mbox{\boldmath ${F}$}}{\mu} \frac{\partial {f}}{\partial
\mbox{\boldmath $v$} } =C({ f})\ \ .
\end{equation}
The external force $\mbox{\boldmath $F$}$, acting on the generic
monad, is supposed to derive from a scalar potential,
$\mbox{\boldmath $F$}=-\partial {\cal V}/\partial \mbox{\boldmath
$x$}$. However the theory developed here can be easily generalized
in order to treat particle systems immersed in an electro-magnetic
field \cite{GK}. We don't make any assumption about the nature of
the interaction between the monads except that, during the
collisions the monad number, momentum and energy are conserved.
This assumption implies that the three functions
$g_1(\mbox{\boldmath $v$})=1$, $g_2(\mbox{\boldmath
$v$})=\mbox{\boldmath $v$}$ and $g_3(\mbox{\boldmath
$v$})=\mbox{\boldmath $v$}^2$ are the collisional invariants of
the system and thus the collisional integral $C(f)$ satisfies the
conditions \cite{L,Gla}:
\begin{equation}
\int C(f) d^{n} v= \int \mbox{\boldmath $v$} C(f) d^{n} v= \int
\mbox{\boldmath $v$}^{2}C(f) d^{n} v=0 \ \ .
\end{equation}
In the following, we indicate with ${\cal G}(t, \mbox{\boldmath
$x$} , \mbox{\boldmath $v$})$ the density of a given physical
quantity so that ${\cal G} f d^{n}v d^{n}x$ represents its value
around the point $(\mbox{\boldmath $x$},\mbox{\boldmath $v$})$
while its total value is $\int{\cal G} f d^{n}vd^{n}x$. We
consider now the projection of the system dynamics in the
$n$-dimensional physical space, where the distribution function is
$\rho(t, \mbox{\boldmath $x$})= \int f d^{n} v $. The mean value
of ${\cal G}(t, \mbox{\boldmath $x$} , \mbox{\boldmath $v$})$ in
the point $ \mbox{\boldmath $x$}$ is given by:
\begin{equation}
<{\cal G}(t, \mbox{\boldmath $x$} , \mbox{\boldmath $v$})>
\!\!_{v} =\frac{\int {\cal G}(t, \mbox{\boldmath $x$} ,
\mbox{\boldmath $v$}) f(t, \mbox{\boldmath $x$} , \mbox{\boldmath
$v$} ) d^{n}v} {\int f(t, \mbox{\boldmath $x$} , \mbox{\boldmath
$v$} )
 d^{n} v} \ \ ,
\end{equation}
and represents  the density of the physical quantity in the
physical space. Consequently, its total value can be written as
$\int\rho<{\cal G}>\!\!_{v}d^nx$. We define the densities in the
physical space of some quantities  used in following. The density
of current is given by  $\mbox{\boldmath $u$} =$ $<\mbox{\boldmath
$v$}>\!\!_{v}$. The density of stress tensor is given by
\begin{equation}
\sigma_{ij}=\mu<(v_i-u_i)(v_j-u_j)> \!\!_{v} \ \ ,
\end{equation}
and is  a symmetric tensor $ \sigma_{ij} = \sigma_{ji} $ of rank
two. The density of the heat flux vector is given by
\begin{equation}
h_i =\frac{1}{2}\mu<| \mbox{\boldmath $v$}- \mbox{\boldmath
$u$}|^2 (v_i-u_i)
>\!\!_{v} \ \ .
\end{equation}
The density of energy is given by
\begin{eqnarray}
E=\frac{1}{2}\mu<\mbox{\boldmath $v$}^2>\!\!_{v}+ {\cal V }
=\frac{1}{2}\mu \mbox{\boldmath $u$}^2 + \varepsilon + {\cal V} \
\ ,
\end{eqnarray}
where $ \varepsilon$ is the density of the internal energy defined
as:
\begin{equation}
\varepsilon=\frac{1}{2}\mu \left (<\mbox{\boldmath
$v$}^2>\!\!_{v}-
 <\mbox{\boldmath $v$}>\!\!_{v}^2 \right ) \ \ .
\end{equation}
It results that $\varepsilon=\sigma_{ii}/2$ and the pressure is
defined as $\pi=\rho \, \sigma_{ii}/n$. Then we can write the
equation of state for the system as $\pi=2\,\rho \, \varepsilon/n
$.

Multiplying Eq. (2) by the three collisional invariants  and after
integration with respect to $v$, the three following hydrodynamic
equations can be obtained \cite{L}:
\begin{equation}
\frac{\partial \rho}{\partial t}
+\frac{\partial }{\partial \mbox{\boldmath $x$}}
(\rho\mbox{\boldmath $u$})=0\ \ ,
\end{equation}
\begin{equation}
\frac{\partial}{\partial t}(\mu\rho u_i) + \frac{\partial
}{\partial x_j}(\rho \varphi_{ij}) +\rho\frac{\partial {\cal V
}}{\partial x_i}=0 \ \ ,
\end{equation}
\begin{equation}
\frac{\partial}{\partial t}(\rho E) +\frac{\partial }{\partial
\mbox{\boldmath $x$}} (\rho\mbox{\boldmath $s$})
-\rho\frac{\partial {\cal V}}{\partial t}=0\ \ ,
\end{equation}
where $\varphi_{ij}=\mu u_iu_j+\sigma_{ij}$ is the momentum flux
density tensor and $s_i= E u_i +  \sigma_{ij}u_j +h_i$ is the
energy flux density vector.

Eq. (9) is the continuity equation for the system which behaves in
the physical space as a fluid. This equation implies the
conservation  of the particle number:
\begin{equation}
N=\int \rho \, d^{n} x \ \ .
\end{equation}
Eq. (10) is the Euler equation for the fluid and in absence of
external forces $\partial {\cal V}/\partial \mbox{\boldmath
$x$}=0$, implies the conservation of the system total momentum:
\begin{equation}
\mbox{\boldmath $P$}=\int \mu \,\rho\,\mbox{\boldmath $u$}\, d^{n}
x \ \ .
\end{equation}
Finally, Eq.(11) governs the evolution of energy and in the case
of time independent forces $\partial {\cal V}/\partial t=0$
implies the conservation of the system total energy:
\begin{equation}
H=\int E\, \rho\, d^{n} x \ \ .
\end{equation}

By taking into account Eq. (9) we rewrite Eq. (10) in the
following form
\begin{eqnarray}
\mu\frac{D \mbox{\boldmath $u$} }{D t}=  \mbox{\boldmath ${\cal
F}$}^{(\sigma)}+ \mbox{\boldmath $F$} \ \ ,
\end{eqnarray}
where $D/Dt=\partial/\partial t + \mbox{\boldmath $u$}\cdot\,
\partial/\partial \mbox{\boldmath
$x$}$ is the total time derivative while the hydrodynamic force
$\mbox{\boldmath ${\cal F}$}^{(\sigma)}$, is originated from the
stress tensor and takes the form
\begin{equation}
{\cal F}^{(\sigma)}_j= -\sigma_{jk}\frac{\partial \xi}{\partial
x_k}-\frac{\partial \sigma_{jk}}{\partial x_k} \ \ ,
\end{equation}
 with
\begin{equation}
{\xi} = \ln \rho \ \ .
\end{equation}

It is interesting to note that the force $-\mu D \mbox{\boldmath
$u$}/Dt$ assumes a Lorentz like form
\begin{equation}
- \mu\frac{D \mbox{\boldmath $u$} }{D t} =\mu \mbox{\boldmath
${\cal E}$}^{(u)}+\mu \mbox{\boldmath $u$}\times\mbox{\boldmath
${\cal B}$}^{(u)} \ \ ,
\end{equation}
where the fields $\mbox{\boldmath ${\cal E}$}^{(u)}$ and
$\mbox{\boldmath${\cal B}$}^{(u)}$ can be derived from the four
vector potential
\begin{equation}
{\cal A}^{^{\scriptstyle(u)}}=\left({\cal
A}_0^{^{\scriptstyle(u)}},\mbox{\boldmath${\cal
A}$}^{^{\scriptstyle(u)}}\right)=\left(\frac{1}{2}\mbox{\boldmath$u$}^2,
\mbox{\boldmath$u$}\right) \ \ .
\end{equation}
by means of
\begin{equation}
\mbox{\boldmath ${\cal E}$}^{^{\scriptstyle(u)}} =-\frac{\partial
{\cal A}_0^{^{\scriptstyle(u)}}}{\partial \mbox{\boldmath
$x$}}-\frac{\partial \mbox{\boldmath${\cal
A}$}^{^{\scriptstyle(u)}}}{\partial t} \ \ ,
\end{equation}
\begin{equation}
\mbox{\boldmath ${\cal B}$}^{^{\scriptstyle(u)}}
=\frac{\partial}{\partial\mbox{\boldmath $x$}}
\times\mbox{\boldmath${\cal A}$}^{^{\scriptstyle(u)}} \ \ .
\end{equation}
The structural similarity between $-\mu D \mbox{\boldmath $u$}/Dt$
and the Lorentz force it is important when we treat systems in the
presence of external electromagnetic fields \cite{GK}.

\section{Derivation of quantum potential}

The absence of the magnetic term in the external force
$\mbox{\boldmath $F$}$ permits us to assume that the particle
system is spinless  \cite{T1,T2,KK}. Then we can set
$\mbox{\boldmath ${\cal B}$}^{(u)}=0$ and consequently we have
that $\mbox{\boldmath$u$}$ is irrotational,
$(\partial/\partial\mbox{\boldmath $x$}) \times \mbox{\boldmath
${u}$}=0$, and can be derived from a scalar potential
\begin{equation}
\mbox{\boldmath $u$}=\frac{1}{\mu}\frac{\partial S}{\partial
\mbox{\boldmath $x$}}\ \ .
\end{equation}
 At this point we observe that all the terms in  Eq.(15)
are irrotational. In particular also the stress force
$\mbox{\boldmath ${\cal F}$}^{(\sigma)}$ is irrotational,
$(\partial/\partial\mbox{\boldmath $x$}) \times  \mbox{\boldmath
${\cal F}$}^{(\sigma)}=0$, and it can be derived from a scalar
potential,
\begin{equation}
\mbox{\boldmath ${\cal F}$}^{(\sigma)} =-\frac{\partial {\cal W
}}{\partial\mbox{\boldmath{$x$}}} \ \ ,
\end{equation}
 so that for
spinless systems we have the following condition:
\begin{equation}
\sigma_{jk} \frac{\partial {\xi}}{\partial x_k}+ \frac{\partial
\sigma_{jk} }{\partial x_k} = \frac{\partial {\cal W}}{\partial
x_j} \ \ .
\end{equation}
The vectorial equation (15) becomes now a scalar equation:
\begin{equation}
\frac{\partial S}{\partial t} +\frac{1}{2\mu}\left (\frac{\partial
S}{\partial \mbox{\boldmath $x$}} \right ) ^2 + {\cal W} + {\cal
V} = 0 \ \ .
\end{equation}

We remark that Eq. (24) allows us to link the quantities
$\sigma_{jk}$ and ${\cal W}$. A first solution of Eq. (24),
describing a classical Eulerian fluid, is given by
$\sigma_{jk}=\frac{2}{n}\,\varepsilon(\xi)\delta_{jk}$ and ${\cal
W}=\frac{2}{n}\,\varepsilon(\xi)+ \frac{2}{n}\int \varepsilon(\xi)
d\xi$ where the internal energy $\varepsilon(\xi)$ is an arbitrary
algebraic function.

In the following, we will show that Eq.(24) admits a second more
interesting and less evident solution. If this solution exists
both terms in the left hand side of (24) will take the form
$\partial(...)/\partial x_j$. In particular, for the second term
we must pose $\sigma_{jk}=\partial a_k/\partial x_j$. Using the
symmetry of the density of the stress tensor
$\sigma_{jk}=\sigma_{kj}$ imposed by Eq.(5), we have that
$a_k=\partial \alpha/\partial x_k$. Then the density of the stress
tensor assumes the form
\begin{equation}
\sigma_{jk}=\frac{\partial^2 \alpha}{\partial x_j\partial x_k }\ \
,
\end{equation}
 with $\alpha$ an unknown scalar function depending
on the field $\xi$. Eq. (24) can be written now in the form
\begin{equation}
\frac{\partial^2 \alpha }{\partial x_j\partial x_k} \,
\frac{\partial {\xi}}{\partial x_k} = \frac{\partial}{\partial
x_j}\left( {\cal W}-\frac{\partial^2 \alpha }{\partial x_k\partial
x_k}\right) \ \ .
\end{equation}
By making the hypothesis that the function $\alpha$ is an
algebraic function of the field ${\xi}$ (it can be easily verified
that this is the only possibility), the left hand side in (27),
after developing the derivatives, becomes
\begin{eqnarray}
\frac{\partial^2 \alpha }{\partial x_j\partial x_k} \,
\frac{\partial {\xi}}{\partial x_k} =\frac{\partial}{\partial
x_j}\left[ \frac{1}{2} \frac{d {\alpha}}{d{\xi}}\left(\!
\frac{\partial {\xi}}{\partial x_k}\!\right)^{\!\!2} \right] \!
\!+ \!\! \frac{1}{2} \frac{d^2 {\alpha}}{d{\xi}^2}
 \frac{\partial {\xi}}{\partial x_j}
 \left(\! \frac{\partial {\xi}}{\partial
x_k}\!\right)^{\!\!2} \nonumber
\end{eqnarray}
and must be written in the form $\partial(...)/\partial x_j$. This
condition requires that
\begin{equation}
\frac{d^{\,2} \alpha }{d {\xi}^2}=0 \ \ .
\end{equation}
After integration of this last equation we obtain $\alpha=c{\xi}+
c_o$ with $c$ and $c_o$ arbitrary constants. The constant $c_o$,
not influencing the value of $\sigma_{jk}$, can be set equal to
zero. Consequently, we obtain
\begin{equation}
\sigma_{jk}= -\frac{\eta ^2}{4\mu}\frac{\partial^2 {\xi}}{\partial
x_j
\partial x_k} \ \ ,
\end{equation}
\begin{equation} \varepsilon=-\frac{\eta ^2}{8\mu}
\frac{\partial^2 {\xi} }{\partial \mbox{\boldmath $x$}^2} \ \ ,
\end{equation}
where we have posed $c= -\eta ^2/4\mu$ in order to have
$\varepsilon>0$. The integration constant $\eta \geq 0$ of Eq.
(28) is a real free parameter of the theory.

The potential ${\cal W}$ can be calculated immediately from (27)
by posing $\alpha=-(\eta ^2/4\mu)\xi$:
\begin{equation}
{\cal W}=- \frac{\eta ^2}{4\mu} \left[ \frac{1}{2} \left(
\frac{\partial {\xi} } {\partial \mbox{\boldmath $x$}} \right)^2
+\frac{\partial^2 {\xi} } {\partial \mbox{\boldmath $x$}^2}\right
] \ \ .
\end{equation}
Equations (29)and (31) describe a fluid different from the
Eulerian one previously obtained. We call it quantum fluid, being
${\cal W}$ the Mandelug-Bohm quantum potential. This fluid will be
the object of our study in the following. We remark that the
procedure used here to obtain the quantum potential evidences
clearly its origin. In order to obtain (31) we have used the
definition (16) of $\mbox{\boldmath ${\cal F}$}^{(\sigma)}$, the
property $\sigma_{jk}=\sigma_{kj}$ enforced by the definition (5)
of $\sigma_{jk}$ and additionally assuming that $\mbox{\boldmath
${\cal F}$}^{(\sigma)}$ is a conservative force, being the system
a spinless one. Besides, the fundamental constant $\hbar=N\eta$
emerges naturally as the integration constant of (28) and is a
free parameter for the theory.

Finally we note that the expression of the density of stress
tensor $\sigma_{jk}$ given by Eq. (29), in the literature
\cite{T1,CW}, appears in a different form, namely:
\begin{eqnarray}
\sigma_{jk}= -\frac{\eta
^2}{4\mu}\left(\frac{1}{\rho}\frac{\partial^2 \rho }{\partial
 \mbox{\boldmath $x$}^2}\delta_{jk} -\frac{1}{\rho ^2}
  \frac{\partial \rho}{\partial x_j}
\frac{\partial \rho}{\partial x_k}\right).  \nonumber
\end{eqnarray}
Unfortunately, by considering the inverse procedure {\it
Shr\"odinger equation} $\rightarrow$ {\it quantum fluid equations}
adopted in the literature, both the expressions can be obtained,
but no clear criterium exists to decide which of them is correct.

\section{Schr\"odinger equation}

The quantum fluid is described completely by the two scalar fields
$\rho$, $S$ whose evolution equations are (9) and (25),
respectively. Alternatively, we can describe this fluid by means
of the two scalar fields ${\xi}$, $S$ whose evolution equations
are:
\begin{equation}
\frac{\partial {\xi}}{\partial t} +\frac{1}{\mu} \frac{\partial^2
S }{\partial \mbox{\boldmath $x$}^2} +\frac{1}{\mu}\frac{\partial
{\xi}}{\partial \mbox{\boldmath $x$}}
 \frac{\partial S}{\partial \mbox{\boldmath $x$}}=0 \ \ ,
\end{equation}
\begin{equation}
\frac{\partial S}{\partial t} -\frac{\eta ^2}{4\mu}
\frac{\partial^2 {\xi}}{\partial \mbox{\boldmath $x$}^2}
-\frac{\eta ^{\!2}}{8\mu}\left (\!\frac{\partial {\xi}}{\partial
\mbox{\boldmath $x$}}\!\right ) ^{\!2} \!\!+
\!\frac{1}{2\mu}\!\left (\frac{\partial S}{\partial
\mbox{\boldmath $x$}}\! \right ) ^2\!\! + {\cal V}= 0 \ \ .
\end{equation}
Given the structural resemblance of (32) and (33), we introduce
the complex scalar field $\Omega$ through:
\begin{equation}
\Omega=\frac{\xi}{2}+\frac{i}{\eta}S \ \ .
\end{equation}
This field satisfies the condition $\int |\exp \Omega|^2 d^{n}
x=N$ and is a many-valued function analogously to the function
$S$, which is defined by means of (22). We can describe now the
quantum fluid by using the complex field $\Omega$ which obeys the
evolution equation:
\begin{equation}
i\eta\frac{\partial \Omega}{\partial t}= -\frac{\eta^2}{2\mu}
\left[ \frac{\partial^2 \Omega}{\partial\mbox{\boldmath $x$}^2}
+\left(\frac{\partial \Omega}{\partial\mbox{\boldmath
$x$}}\right)^2 \right] +{\cal V}  \ \ .
\end{equation}
Eq. (35) is a Burger like equation which can be linearized by the
Hopf-Cole transformation
\begin{equation}
\Omega={\rm Log} \Psi \ \ ,
\end{equation}
with ${\rm Log}$ representing the many-valued complex logarithmic
function. If we introduce the total mass $m=N\mu$, the potential
$V=N{\cal V}$ under which the entire system evolves and after
setting $\hbar=N\eta$, Eq.(35) transforms to the Schr\"odinger
equation:
\begin{equation}
i\hbar\frac{\partial \Psi}{\partial t}= -\frac{\hbar^2}{2m}
\frac{\partial^2\Psi}{\partial\mbox{\boldmath $x$}^2}+ V \Psi
 \ \ .
\end{equation}
The field $\Psi$ satisfies the normalization condition
\begin{equation}
\int |\Psi|^2 d^{n} x=N \ \ ,
\end{equation}
and is a single-valued function. This last restriction implies
that $\Omega$ is a many-valued logarithm so that $S/\eta$ is a
many-valued function, whose different values differ by integer
multiples of $2\pi$. This restriction on the values of $S$ implies
\begin{equation}
\oint_\gamma \mbox{\boldmath $u$}d\mbox{\boldmath
$x$}=j2\pi\hbar/m \ \ ,
\end{equation}
and represents the quantization condition for the system
\cite{KK,TW}.

The system of mass $m$, namely the quantum particle, can be
described by means of the field $\Psi$ taking into account that
$|\Psi|^2d^nx$ represents the number of monads around the point
$\mbox{\boldmath $x$}$ or alternatively, in probabilistic terms,
by means of the field $\psi=\Psi/\sqrt{N}$ according to the
Copenhagen School. In the following we will continue to describe
the system by using the field $\Psi$. The ansatz
\begin{equation}
\Psi=\rho^{1/2}\exp \left(\frac{i}{\hbar}{\rm S}\right)\ \ ,
\end{equation} with
${\rm S}=NS$, appears now as the transformation which linearizes
the evolution equations (32), (33) and at the same time selects
the quantized states of the system.

\section{Quantum operators}

We show now that the statistical nature of the quantum particle
permits us to introduce the quantum operators whose mean values
give the measured values of the physical observables. We recall
that the measured value for a given physical observable of a
statistical system is the mean value of its associated density
${\cal G}(t, \mbox{\boldmath $x$} , \mbox{\boldmath $v$})$ and can
be calculated starting from the distribution function $f$:
\begin{equation}
<{\cal G}(t, \mbox{\boldmath $x$} , \mbox{\boldmath $v$})>
\!\!_{vx} =\frac{\int {\cal G}(t, \mbox{\boldmath $x$} ,
\mbox{\boldmath $v$}) f(t, \mbox{\boldmath $x$} , \mbox{\boldmath
$v$} ) d^{n}v d^{n} x } {\int f(t, \mbox{\boldmath $x$} ,
\mbox{\boldmath $v$} )
 d^{n} v d^{n} x} \ \ .
\end{equation}
In the case of the quantum fluid we can calculate the above mean
value also starting from $\Psi$. In fact if we consider the
identities
\begin{eqnarray}
 \int\mbox{\boldmath $v$}^l\,f\, d^{n} v\, d^{n}x
  =\int d^{n}x \, \Psi^*\left (\frac{-i\hbar}{m} \frac{\partial}{\partial
\mbox{\boldmath $x$}}\right )^l \Psi   \ \ ,
\end{eqnarray}
for $l=1$ and $l=2$ (see appendix) it is easy to verify that the
physical observables momentum and kinetic energy  can be obtained
also as mean values of the two operators
\begin{equation}
\widehat{\cal \mbox{\boldmath $p$}}
=-i\hbar\frac{\partial}{\partial \mbox{\boldmath $x$}}\ \ ,
\end{equation}
and
\begin{equation}
\widehat{T} =\frac{\widehat{\cal \mbox{\boldmath $p$}}^{\,2}}{2m}
\ \ ,
\end{equation}
respectively, having defined the mean value of an operator as:
\begin{equation}
<{\widehat{\cal G}}> \!\!_{x}
 =\frac{\int \Psi^*(t, \mbox{\boldmath
$x$}){\widehat{\cal G}}\Psi(t, \mbox{\boldmath $x$}) d^{n}x}
 {\int \Psi^*(t, \mbox{\boldmath
$x$})\Psi(t, \mbox{\boldmath $x$})
 d^{n} x} \ \ .
\end{equation}
When the density of the physical observable is an arbitrary
function of the type ${\cal G}(t, \mbox{\boldmath $x$})$ it is
immediate to verify that $\widehat{\cal G}= {\cal G}(t,
\mbox{\boldmath $x$})$ and then  we have
\begin{equation}
<\!{\cal G}(t, \mbox{\boldmath $x$})\!> \!\!_{vx}=$ $<\!{\cal
G}(t, \mbox{\boldmath $x$})\!> \!\!_{x}=<{\widehat{\cal G}}>\!\!_x
\ \ .
\end{equation}
We remark at this point the statistical origin of the quantum
operators. In fact we have introduced the quantum operators using
the definition (41) of the mean value of the physical observables
of a many body classical system  and obviously the properties of
the quantum fluid, namely the expressions of $\mbox{\boldmath
$u$}$ and $\varepsilon$ (see appendix).

In the following, we will write the total energy $H$ of the
quantum fluid in terms of the field $\Psi$. For this reason, we
observe that any function ${\cal G}(t, \mbox{\boldmath $x$})$ can
be written as
\begin{equation}
{\cal G}={\cal G}^{(0)}+{\cal G}^{(I)} \ \ ,
\end{equation}
where the first part has mean value equal to zero
\begin{equation}
<\!{\cal G}^{(0)}\!>\!\!_{x} =0 \ \ ,
\end{equation}
so that $<{\cal G}>\!\!_{x} =$ $ <{\cal G}^{(I)}>\!\!_{x}$. It is
easy to verify that:
\begin{equation}
\varepsilon^{(I)}= {\cal W}^{(I)}= \frac{\eta^2}{8\mu} \left(
\frac{\partial {\xi}}{\partial \mbox{\boldmath $x$}} \right )^2 \
\ .
\end{equation}
The total internal energy ${\cal H}=\int \varepsilon \rho
\,\,d^{n} x$ of the system is given by
\begin{equation}
{\cal H}=\frac{\eta^2}{8\mu }\int \left( \frac{\partial
{\xi}}{\partial \mbox{\boldmath $x$}} \right )^2 \rho \,\,d^{n} x
\ \ ,
\end{equation}
while the total energy $H$ given by
\begin{equation}
H=\int  \left (\frac{1}{2}\mu\mbox{\boldmath $v$} ^2
 + {\cal V} \right)f d^{n} v d^{n} x \ \ .
\end{equation}
assumes the form
\begin{equation}
H=\int \left [\frac{1}{2\mu}\left (\frac{\partial S} {\partial
\mbox{\boldmath $x$}}\right ) ^2 + \frac{\eta^2}{8\mu} \left(
\frac{\partial {\xi}}{\partial \mbox{\boldmath $x$}} \right )^2 +
{\cal V} \right ] \rho \, d^{n} x \ \ ,
\end{equation}
and is the sum of two terms:
\begin{equation}
H=H_{cl}+ {\cal H} \ \ ,
\end{equation}
with
\begin{equation}
H_{cl}=\int \left [\frac{1}{2\mu}\left (\frac{\partial S}
{\partial \mbox{\boldmath $x$}}\right ) ^2  + {\cal V} \right ]
\rho \ d^{n} x \ \ .
\end{equation}
The classical term $H_{cl}$ of the total energy  corresponds to a
system with $\sigma_{ij}=0$ while the quantum term ${\cal H}$ is
originated by the internal structure of the system and is simply
its {\it internal energy}.

\noindent The total energy can be written in the form
\begin{equation}
H= \int{ \left (\frac{\eta ^2}{2\mu}\left | \frac{\partial \Omega}
{\partial \mbox{\boldmath $x$}} \right | ^2 + {\cal V} \right
)\rho} \ d^{n} x \ \ .
\end{equation}
while in terms of $\Psi$ becomes
\begin{equation}
H= \int{ \left (\frac{\eta ^2}{2\mu}\left | \frac{\partial \Psi}
{\partial \mbox{\boldmath $x$}} \right | ^2 + {\cal V}\, |\Psi|^2
\right )} d^{n} x \ \ .
\end{equation}
When the system is described by means of the field $\Psi$ is
canonic  and the evolution equation (37) can be obtained starting
from the Hamiltonian (56) by using a variational principle:
\begin{equation}
i\eta \frac{\partial \Psi}{\partial t}= \frac{\delta H }{\delta
\Psi^*} \ \ .
\end{equation}
Thus the ansatz $\Psi=\rho^{1/2}\exp (i{ S}/\eta)$  can be
viewed as a transformation which linearizes the evolution equation
and at the same time conserves the canonicity of the system.

\section{Uncertainty Principle}

The internal structure of the quantum particle implies a spatial
dispersion and consequently an indetermination
\begin{equation}
(\Delta x)^2=<\!(\mbox {\boldmath $x$}- <\!\!\mbox{\boldmath $x$}
\!\!>\!\!_{x})^2\!>\!\!_{x} \ \ ,
\end{equation}
on the measure of his position. In order to relate $\Delta x$ with
${\cal H}$ we consider the inequality
\begin{equation}
\int \left|(<\!\!\mbox{\boldmath $x$}\!\!>_x \!\!- \mbox{\boldmath
$x$})\, \frac{\partial \rho}{\partial \mbox{\boldmath $x$}}\right|
d^{n}x \geq N \ \ ,
\end{equation}
which can be immediately verified if we observe that it becomes an
obvious equality when the integrand is considered without the
absolute value. This inequality can be written also as:
\begin{equation}
\int \sqrt{\frac{1}{N}\left(\mbox{\boldmath$x$}-<\!\mbox{\boldmath
$x$}\!\!>_x\right)^2\rho } \,\,  \sqrt{\frac{1}{N} \left
(\frac{\partial {\xi}}{\partial \mbox{\boldmath $x$}}\right)^2\!\!
\rho}
 \,\, d^{n}x \geq 1 \ \ ,
\end{equation}
which, by taking into account the Schwartz inequality assumes the
form
\begin{equation}
\frac{1}{N}\!\int \left(\mbox{\boldmath$x$}-<\!\mbox{\boldmath
$x$}\!\!>_x\right)^2\rho
 \,\, d^{n}x \,\,   \frac{1}{N}\!\int \left (\frac{\partial {\xi}}{
\partial \mbox{\boldmath $x$}}\right)^2 \!\rho \,\, d^{n}x
 \geq 1 \ .
\end{equation}
After remembering the definitions of $\Delta x$ and ${\cal H}$ the
inequality  (61), transforms as
\begin{equation}
\Delta x \, \sqrt{2m{\cal H}}\geq \frac{\hbar}{2} \ \ .
\end{equation}
We show now that ${\cal H}$ is related also with the uncertainty
on the measure of the momentum  of the system
\begin{eqnarray}
(\Delta p)^2&&=\!m^2\!\!<\!\! (\mbox{\boldmath $v$}-
<\!\!\mbox{\boldmath $v$}\!\!>\!\!_{vx})^2\!\!>\!\!_{vx}\nonumber
\\ &&=<\!\!(\mbox{\boldmath $\widehat{p}$}- <\!\!\mbox{\boldmath
$\widehat{p}$}\!>\!\!_{x})^2\!\!>\!\!_{x} \ \ .
\end{eqnarray}
In fact by taking into account that ${\cal H}$ can be written also
in the form
\begin{equation}{\cal H}=\frac{1}{2}m \!\!<\!(\mbox
{\boldmath $v$}- <\!\mbox{\boldmath $v$}
\!>\!\!_{v})^2\!\!>\!\!_{vx} \ \ ,
\end{equation}
and after some simple algebra we obtain the relation:
\begin{equation}
(\Delta p)^2=(\Delta p_{cl})^2+2m{\cal H} \ \ \ ,
\end{equation}
where
\begin{equation}
(\Delta p_{cl})^2=m^2\!<\!(\mbox {\boldmath $u$}-
<\!\mbox{\boldmath $u$}\!
>\!\!_{x})^2\!\!>\!\!_{x}\ \ .
\end{equation}
Then from (65) we conclude
\begin{equation}
\Delta p\! \geq\! \sqrt{2m{\cal H}} \ \ .
\end{equation}
This last inequality if combined with (62) leads immediately to
the Heisenberg uncertainty principle:
\begin{equation}
\Delta x \,\, \Delta p \geq \frac{\hbar}{2} \ \ .
\end{equation}
The procedure used to obtain this principle shows clearly its
statistical origin.

\section{Concluding remarks}

The theory here developed can be viewed as an approach
constituting a deductive and direct method which, in a
self-consistent scheme of a classical many-body physics, permits
us i) to obtain the probabilistic nature of the quantum
description and to interpret the wave function $\psi$ according to
the Copenhagen school; ii) to derive naturally the quantum
potential and then the Schr\"odinger equation; iii) to calculate
the values of the physical observables as mean values of certain
associated operators, namely the quantum operators; iv) to obtain
the Heisenberg uncertainty principle. Finally the fundamental
constant $\hbar$ emerges naturally as an integration constant and
represents a free parameter for the theory.

The theory can be viewed also as describing a mechanism allowing
us to build the quantum particle starting from its constituents.
The quantum particle of mass $m$ turns out to be a statistical
system having a spatial extension and an internal structure. It is
composed by $N$ identical subquantum interacting particles of mass
$\mu$, the monads. These monads obey the laws of classical physics
and their dynamics is described in the phase space by the standard
kinetic equation. We don't make any assumption about the nature of
the interaction between the monads. We assume only that during
collisions the monad number, momentum and energy are conserved.
These assumptions, in the case of a spinless system, imply that
its dynamics in the physical space is governed by the
Schr\"odinger equation.

It is now clear that quantum mechanics is a non local, hidden
variables theory, as suspected by some of its founders. In fact
Eq. (2) describes a non relativistic subquantum statistical
ensemble. The hidden dynamics, in which Einstein believed, seems
to be the one imposed by the subquantum {\it monadic kinetics}.
The orthodox quantum mechanics is an axiomatic theory and then, by
taking into account the G\"odel theorem, it can be obtained only
in the framework of a wider metaquantum theory. This metaquantum
theory appears to be now simply the subquantum  monadic kinetics.

We discuss now briefly the problem concerning the locality in
quantum physics, that were left unresolved in the twenty three
year long debate between Einstein and Bohr and were reconsidered
by Bell in 1964. It is well known that the Bell's inequality has
been obtained in the framework of local, hidden variables and
deterministic theories. This inequality is in disagreement both
with quantum mechanics and experimental evidence. The reason of
this disagreement now appears clear. Here we have obtained the
quantum mechanics staring from the underlying monadic kinetics
which is a non local, hidden variables and probabilistic theory.
At this point, spontaneously arises the question, if it is
possible to include the locality in quantum physics. After noting
that a relativistic kinetics is a local and probabilistic theory,
we can make the conjecture that {\it a subquantum relativistic
monadic kinetics could be underlying a local quantum theory}.

\appendix
\section{}

In this appendix we demonstrate the identity (42) in the case
$l=2$. We take into account the definition of the internal energy
(8) and the expression of $u_i$ and $\varepsilon$ given by (22)
and (30) respectively, we have:
\begin{eqnarray}
\int v_i^2fd^nxd^nv&=&\int d^nx \rho<v_i^2>_v=\int d^nx
\rho(u_i^2+2\mu\varepsilon)  \nonumber \\ &=& \int \rho
\left[\left( \frac{1}{\mu}\frac{\partial S}{\partial x_i}
\right)^2-\frac{\eta^2}{4\mu^2}\frac{\partial^2 \ln \rho}{\partial
x_i^2 }\right]d^nx \nonumber \\ &=& -\frac{\eta^2}{\mu^2}\int
\rho\bigg[ \frac{\partial}{\partial x_i}\left(
\frac{1}{2\rho}\frac{\partial \rho}{\partial
x_i}+\frac{i}{\eta}\frac{\partial S}{\partial x_i}
\right)\nonumber
\\ &+&\left( \frac{1}{2\rho}\frac{\partial \rho}{\partial
x_i}+\frac{i}{\eta}\frac{\partial S}{\partial x_i} \right)^2
\bigg] d^nx \nonumber \\ &=& -\frac{\eta^2}{\mu^2}\int
\Psi^*\bigg[\Psi \frac{\partial}{\partial x_i}\left(
\frac{1}{2\rho}\frac{\partial \rho}{\partial
x_i}+\frac{i}{\eta}\frac{\partial S}{\partial x_i} \right)
\nonumber
\\ &+&\left( \frac{1}{2\rho}\frac{\partial \rho}{\partial
x_i}+\frac{i}{\eta}\frac{\partial S}{\partial x_i} \right)^2\Psi
\bigg]d^nx  \nonumber \ \ .
\end{eqnarray}
From the ansatz $\Psi=\rho^{1/2}\exp (iS/\eta)$ we have
\begin{eqnarray}
\frac{\partial \Psi}{\partial x_i}= \left(
\frac{1}{2\rho}\frac{\partial \rho}{\partial x_i}+ \frac{i}{\eta}
\frac{\partial S}{\partial x_i} \right) \Psi \ \ , \nonumber
\end{eqnarray}
so that Eq. (42) immediately is obtained

\begin{eqnarray}
 &&\!\!\!\!\!\! \int v_i^2fd^nxd^nv \nonumber \\
 &=& -\frac{\eta^2}{\mu^2}\int
 \Psi^*\bigg[\Psi \frac{\partial}{\partial x_i}\left(
\frac{1}{2\rho}\frac{\partial \rho}{\partial
x_i}+\frac{i}{\eta}\frac{\partial S}{\partial x_i} \right)
\nonumber
\\ &+&\left( \frac{1}{2\rho}\frac{\partial \rho}{\partial
x_i}+\frac{i}{\eta}\frac{\partial S}{\partial x_i}
\right)\frac{\partial \Psi}{\partial x_i} \bigg]d^nx \nonumber \\
&=& -\frac{\eta^2}{\mu^2}\int d^nx \, \Psi^*
\frac{\partial}{\partial x_i}\bigg[\left(
\frac{1}{2\rho}\frac{\partial \rho}{\partial
x_i}+\frac{i}{\eta}\frac{\partial S}{\partial x_i}
\right)\Psi\bigg]\nonumber\\ &=& -\frac{\eta^2}{\mu^2}\int d^nx \,
\Psi^* \frac{\partial}{\partial x_i}\bigg\{ \frac{\partial
}{\partial x_i}\left[ \rho^{1/2}\exp{(iS/\eta)}\right]\bigg\}
\nonumber \\ &=& -\frac{\eta^2}{\mu^2}\int d^nx \, \Psi^*
\frac{\partial}{\partial x_i}\bigg( \frac{\partial }{\partial
x_i}\Psi\bigg) \nonumber \\
 &=& \int d^nx \,
\Psi^* \left(-\frac{i\hbar}{m} \frac{\partial}{\partial x_i}
\right)^2 \Psi  \ . \nonumber
\end{eqnarray}

\end{document}